\documentclass{sf2a-conf2024}
\usepackage{graphicx}
\usepackage{hyperref}
\usepackage{wrapfig}
\usepackage[]{natbib}  
\usepackage{epstopdf}
\usepackage{amsmath}

\def\BibTeX{{\rm B\kern-.05em{\sc i\kern-.025em b}\kern-.08em
    T\kern-.1667em\lower.7ex\hbox{E}\kern-.125emX}}
\bibpunct{(}{)}{;}{a}{}{,}  %%%%%%%%%%%%%  A&A bibliography style
\begin{document}

\TitreGlobal{SF2A 2024}

%%-----------------------------------------------------------------
%%      the top matter
%%
%hrefs can be introduced as follows: \url{http://www.sf2a.eu/}.

\title{SF2A Environmental Transition Commission:\\ What kind of astrophysics research for a sustainable world?}

\runningtitle{Review SF2A-2024 S11}

\author{F.~Cantalloube}\address{Univ. Grenoble Alpes, CNRS, IPAG, F-38000 Grenoble, France}
\author{D.~Barret}\address{IRAP, Université de Toulouse, CNRS, CNES, UT3-PS, Toulouse, France}
\author{M.~Bouffard}\address{Laboratoire de Plan\'etologie et G\'eosciences, CNRS UMR 6112, Nantes Universit\'e, Universit\'e d'Angers, Le Mans Universit\'e,
Nantes, France}
%\author{N.~Flagey}\address{Space Telescope Science Institute, 3700 San Martin Drive, Baltimore, MD, 21218, USA}
\author{P.~Hennebelle}\address{Universit\'e Paris-Saclay, Universit\'e Paris Cit\'e, CEA, CNRS, AIM, 91191, Gif-sur-Yvette, France}
\author{J.~Milli$^1$}
\author{F.~Malbet$^1$}
\author{A.~Santerne}\address{Aix-Marseille Univ, CNRS, CNES, LAM, Marseille, France}
\author{N.~Fargette}\address{The Blackett Laboratory, Department of Physics, Imperial College London, London SW7 2AZ, UK}
\author{S.~Bontemps}\address{Laboratoire d’astrophysique de Bordeaux, Univ. Bordeaux, CNRS, B18N, allée Geoffroy Saint-Hilaire, 33615 Pessac, France}
\author{C.~Moutou$^2$}
\author{A.~Mouinié$^2$}
\author{H.~Méheut}\address{Universit\'e C\^ote d’Azur, Observatoire de la C\^otee d’Azur, CNRS, Laboratoire Lagrange, Nice, France}
\author{A.~Saint-Martin}\address{Centre europ\'een de sociologie et de science politique, CNRS, EHESS, Paris 1}
\author{A.~Hardy}\address{Sciences Po Bordeaux 11 all\'ee Ausone 33607 Pessac CEDEX - France}

%% Keep this line, even if the page will be settled afterwards.
\setcounter{page}{237}

%%-----------------------------------------------------------------

\maketitle

%%-----------------------------------------------------------------
%%        The abstract

\begin{abstract}
During its annual conference in 2024, the French Society of Astronomy \& Astrophysics (SF2A) hosted, for the fourth time, a special session dedicated to discussing the \emph{environmental transition} within the French astrophysics research community. 
This year had a special context: both the CNRS-INSU and the CNES were preparing their scientific perspectives for the period 2025-2030 in the field of Astronomy-Astrophysics (AA). %The half-day workshop was therefore focused on implementing a desirable future for the conduct of astrophysics research in the coming decades, and brought together about 100 participants. 
In this proceeding, we first describe the main actions undertaken by the \emph{Commission Transition Environnementale}. Then, we summarize the discussions held during the half-day workshop, which brought together about 100 participants, and point to forthcoming proceeding, reports and other related resources. A key message is that the French A\&A community is now fully aware that astronomical activities simply cannot thrive indefinitely in the current situation, and seems now eager to seize the opportunity of developing our profession towards a better social and environmental impact.
\end{abstract}

%% Insert the keywords (to appear in the ADS indexing)
%% Keywords must be separated by a comma

%%-----------------------------------------------------------------

\section{Introduction}
%%---------------------
Since 2021, the SF2A annual \emph{Transition Environnementale} workshops have been bringing the French Astronomy and Astrophysics (A\&A) community together to reflect on the means of action that exist, or need to be implemented, to develop our professions, in terms of organisation and methodology, in the face of the impending environmental and social crisis. 
During the 2024 workshop in Marseille (France), the focus was on how A\&A research operates in this context, which naturally places constraints on our working conditions: how to balance our professional role as researchers/teachers and our social role, while respecting the need to change our approach to work through (inter-)national commitments? In this context, the management methods currently in use in higher education and research, in terms of recruitment, assessment and working conditions, do not appear to be viable both at the level of the individual and at the level of research as a whole.

In 2024, the CNRS-INSU and CNES were both in the process of publishing their scientific prospective reports in the field of Astronomy \& Astrophysics. And in both cases, the exercise involved, for the very first time, specific working groups dedicated to the questions of environmental impact: for the CNRS-INSU, the \emph{Transition Carbone et Ecologique} and \emph{T\'elescopes et Territoires} chapters; for the CNES the \emph{Réduction de l’empreinte environnementale des activités scientifiques spatiales} chapter. In this report we summarise the main messages emerging from these working groups, accompanied by the various contributions to the SF2A-S11 special session \emph{Transition Environnementale} that enriched the discussions. 
%In addition to presentations by guest speakers specialising in the subject, we concluded the workshop with a round-table discussion bringing together the various speakers so that everyone can have the opportunity to express their views and put forward a vision that is desirable for everyone.

%%---------------------------------
%%---------------------------------
\section{Actions of the \textit{Commission Transition Environnementale}}
%%-------------------------
%1page
\begin{wrapfigure}{r}{0.2\textwidth}
  \begin{center}
    \vspace{-25pt}
    \includegraphics[scale=0.29]{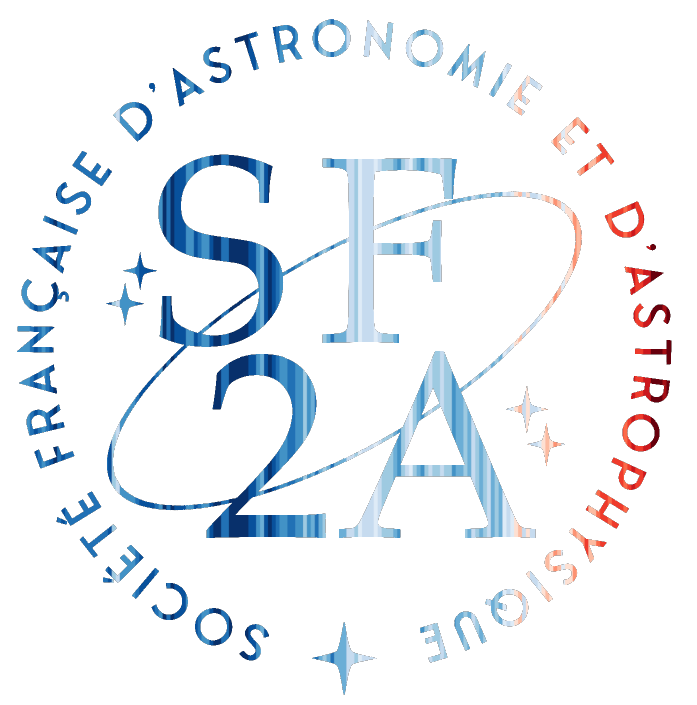}
  \end{center}
  \vspace{-22pt}
  %\caption{SF2A-CTE logo}
\end{wrapfigure}
The \emph{Commission Transition Environnementale} (special Commission for Environmental Transition) of the SF2A, was created following the first dedicated special session at the SF2A 2021 annual conference online.
The aim of the \emph{Commission Transition Environnementale} is to support the French Astronomy \& Astrophysics research community towards a more responsible and resilient way of conducting research in the face of the various environmental crises underway. The commission's activities are grouped around different poles: (a) being a point of contact between researchers and the various research bodies, (b) studies \& analyses, and (c) training, animation \& awareness. 

With the first organizing committee, gathering about 15 French researchers in A\&A from various institutes and working on diverse topics, we wrote in 2022 a letter of intent that explains which actions the \emph{Commission Transition Environnementale} plans to put in place. The \href{https://cloud.univ-grenoble-alpes.fr/s/Roor2REzgWZzXHm}{letter of intent} (in French), aims at ensuring a more virtuous environmental impact of the main missions and activities of the SF2A.

\subsection{Web-page SF2A}
To inform the French A\&A community on the topics of climate change and environmental degradation, we gathered on a single webpage statements, actions and communications led within the SF2A, list of \emph{sustainability working group} point of contact in French A\&A institutes, information about administrative supervising institutes (e.g. CNRS, CNES) and societies (e.g. EAS, IAU), information about collectives of researchers (e.g. Labos1p5, A4E, S4F), and a large number of resources (from publications to raw climate data and tools to raise awareness). All these resources can be found on the \href{https://sf2a.eu/website2023/commission_transition_environmentale/}{\textit{Commission Transition Environnementale}} page of the SF2A website.

\subsection{Survey SF2A-INSU}
The SF2A \emph{Commission Transition Environnementale} and the CNRS-INSU AA prospective working group \emph{Défi climatique et écologique} joined forces to propose a survey based on a 2019 survey led by P.~Martin (IRAP) and whose results were \href{https://sf2a.eu/semaine-sf2a/2019/presentations/S00/Martin_S00.pdf}{presented} during the SF2A annual conference 2019 (Nice). The updated 2024 survey is intended to all research staff, including technical, administrative, and non-permanent. 
In 2019, the survey received about 500 answers within 40 institutes, representing a third of the French A\&A community.
The 2024 version has minor differences compared to the 2019 survey since the goal is to assess the evolution of thoughts during the last 5 years. 
The survey addresses the themes of  (i) research ethics \& culture, (ii) raising awareness \& teaching, (iii) the notion of exemplarity, (iv) individual vs. institutional actions, and (v) career assessment; along four parts (identity, awareness, affirmations and open-boxes). 
Results will be published in the CNRS-INSU AA prospective long documents and by the SF2A (in September 2024, 400 participants contributed).

\subsection{Next steps and actions}
Among the next actions planned by the \emph{Commission Transition Environnementale}, we intend to publish a detailed comparison of the carbon footprint of French A\&A institutes (Santerne et al., in prep.). We are also establishing an environmental charter for the SF2A. % and a set of guidelines to support sustainability within institutes. 
Another project is to propose a survey focusing on early career researchers facing a highly competitive environment, which is not in line with the willingness to reduce one's environmental footprint. %t academic he actions that should be taken to alleviate the climate crises. 

\section{Special sessions during the SF2A annual conference 2024}
As part of the leading actions of the \emph{Commission Transition Environnementale}, we organized a half-day special session dedicated to discussing environmental transition during the SF2A annual conference. %This includes a slot during the morning plenary sessions and a round table during the afternoon session. 
In 2021 (online), the goal was to gauge the community interest to go further into the environmental transition discussions and actions, and we invited V.~Masson-Delmotte for a plenary talk about the 6th assessment report of IPCC. 
In 2022 (Besan\c{c}on), the goal was to assess the carbon footprint of research in astronomy and we invited J.~Knödlseder for a plenary talk about the carbon footprint of research infrastructures for astronomy.
In 2023 (Strasbourg), we discussed the different action levers to reduce our carbon footprint at different levels and we invited E.~Guilyardi to talk about the CNRS ethical committee, on integrating environmental issues into conduct of research as an ethical responsibility. In 2024 (Marseille), we focused on what sustainable future we want for our community.

%In 2024 (Marseille), the topic was focused on what kind of astrophysics research is needed in a sustainable world. 
In the following, we provide a short summary of the discussions led during the 2024 \emph{Commission Transition Environnementale} session. A summary of the 2024 plenary talk can be found in \cite{cantalloube2024a}.

%\subsection{Special sessions 2024}
\paragraph{Prospective CNRS-INSU AA}

The first contributions were focused on the \emph{Astronomie et Sociétés} parts, with chapters \emph{Transition Carbone et Ecologique} (GT~I2, Carbon and Ecological Transition, led by S.~Bontemps) and \emph{T\'elescopes et Territoires, Astronomie Participative} (GT~I3, Telescopes and Territories, Citizen Science, led by C. Moutou). It is noteworthy the first time that working groups on such topics have been put in place. 

The main tasks of the GT~I2, which brought together 10 astrophysicists, were to 
%address the social and institutional requirements by 
(1) summarise existing studies and initiatives to assess and reduce carbon and ecological impacts, (2) propose relevant indicators, policy elements and scenarios for action, while ensuring the support of the community, and (3) suggest measures to support the resilience of the A\&A activities. The chosen methodology was to collect information and lead a number of interviews with various actors of the French A\&A research and gather its analysis in a document to trigger debates and discussions. The \href{https://prospective-aa.sciencesconf.org/data/rapport_synthese_groupeI2_transition_carbone_ecologique.pdf}{GT I2 summary report} is publicly available (in French). 
%\textbf{Une phrase de ccl points cles}.

The tasks of the GT~I3 \emph{T\'elescopes et Territoires, Astronomie Participative}, which brought together 9 astrophysicists from different institutes, were to analyse the situations in which the worlds of professional astronomy and society meet, and to suggest areas for improvement. 
The chosen methodology was to collect information and knowledge from social sciences, to conduct several interviews with representatives of astronomical sites, and to use community surveys in order to assess the level of awareness over our social impacts. The whole process was followed and aided by a sociologist and historian P.~Marichalar (CNRS-IRIS, author of \href{https://www.editionsladecouverte.fr/la_montagne_aux_etoiles-9782348079672}{La montagne aux étoiles}). The \href{https://prospective-aa.sciencesconf.org/data/rapport_synthese_groupeI3_telescopes_territoires_astronomie_participative.pdf}{GT I3 summary report} is publicly available (in French). Many similarities between social and environmental issues can be highlighted, especially in isolated sites.
%\textbf{Une phrase de ccl points cles}.

\paragraph{Prospective CNES AA}
The second part focused on the 2024 CNES science prospective on the future of space research and exploration in France, which has a special group called \emph{Réduction de l'empreinte environnementale des activités scientifiques spatiales} (SG5, Reducing the environmental footprint of space science activities, coordinated by H.~M\'eheut and B.~Millet). Among the artificial satellite fleat, only 1.5\% is dedicated to space science. 
The first step was to assess the carbon footprint of satellites over their full life cycle, which appears to be quite different from one type of satellite to another, but always dominated by manufacturing, Assembly Integrating and Tests, and launch. A second step was to prepare the future trajectories of this space sector, in its social role, pioneering role and responsibilities. The full \href{https://sps2024.com/wp-content/uploads/CNES_RapportSPS2024.pdf}{CNES report} is publicly available (in French).
%\textbf{Une phrase de ccl points cles}.

To complete the view of spatial activities and developments, in the frame of its scientific and social impacts, we invited A.~Saint-Martin, a sociologist specialized in the history of astronautics, to discuss how space is becoming a kind of new Eldorado to be exploited, subject to promises of economic returns with an intensification of space deployment in a logic of massive investment (including start-ups) and competition with other countries. This deeply-rooted culture, however, is at odds with the current discourse, which sometimes ostentatiously displays a space scientific activity that is nonetheless marginal, while evoking the importance of taking ecological constraints into account. 
%Dissonnances actuelles de positionnement. Cette culture tres ancrée est pourtant en dissonance avec les discours actuelles qui montrent de facon parfois ostentatoire une activite scientifique du spatiale pourtant marginale, tout en evoquant l'importance de prendre en compte les contraintes ecologiques. 
More information can be found in the book \href{https://lafabrique.fr/une-histoire-de-la-conquete-spatiale/}{\textit{Une histoire de la conqu\^ete spatiale}}. % (in French)

\paragraph{Early career researchers facing the environmental crises}
A third part was focused on the specific situation of early career researchers (ECR) taken in the highly competitive academic system. The first presentation provided with general considerations about the ecological involvement of early-career researchers, given by M.~Bouffard on behalf of the young researcher multidisciplinary network of the \emph{Labos1point5} collective. Two main points were raised on career-related topics: (1) the evaluation criterion of ECR are not in line with the necessity of reducing one's environmental footprint, and virtuous actions are often neither recognized nor supported; (2) professional instability and lack of job security often enhances the environmental footprint, particularly in the case of expatriation. This second point was specifically supported by evidence shown by N.~Fargette in a second presentation, raising the discussion on who should take responsibility for this carbon footprint: individuals or the institutions that push them to emigrate? This discussion can be found in \cite{fargette2024}. All of these points together create cognitive dissonance, loss of purpose and a sense of loneliness that can significantly alter mental health.  

\paragraph{Research actions: astronomy and ecology}
One new topic engaged during the session was the link between research actions in astronomy and ecology. In the presentation, J.~Milli introduced the actions led by the \emph{Observatoire de l’environnement nocturne à Grenoble} (observatory of the night environment in Grenoble), a structure in which astronomers collaborate with scientists of other domains (environment, ecology,...) about the harmful effects of light pollution. In this case, astronomical instruments are used to increase the data coverage and quantification of light pollution in the Grenoble area. Associated with communication to the general public, such action pushes towards concrete political actions at a local scale. More can be found in \cite{milli2024}. 

\paragraph{Reshaping research in astrophysics in the ecological transition era}
At last, we received A.~Hardy, a PhD student in sociology of science, whose work focuses on the conditions of scientific research in the context of climate change using a qualitative survey methodology. 
%A.~Hardy presented some results of his research in a contributed 
In a talk entitled \emph{Decarbonising astrophysics and the frontiers of scientific work}, he highlighted the levers, barriers and concerns raised during internal discussions within a French astronomy laboratory to implement structural changes that would be consistent with decarbonising the research activities. % The slides (in French) can be found \href{https://cloud.univ-grenoble-alpes.fr/s/oAMDXEwykkgLNFH}{here} and 
Additional work can be found in \cite{hardy2024}. 
%\textbf{Une phrase de ccl points cles}.

This contribution was in line with the presentation given by P.~Hennebelle, revealing some key points collected from interviews with astronomers conducted by the working group \emph{cœur de métier} (core work) in astrophysics of the \emph{Labos1point5} collective. Indeed astrophysics is using large infrastructures such as telescopes \citep[accounting for about 65\% of its carbon footprint;][]{martin2022}, which questions the social utility and the necessary reductions in this research theme \citep{hennebelle2024}.
%\textbf{Une phrase de ccl points cles}.

An example of initiatives carried at the IRAP institute (Toulouse, France) was presented by A.~Mouini\'e, the IRAP environmental transition officer recruited in 2023. Following the development of a detailed carbon footprint, four working groups were set up, focusing on (i) life at the institute, (ii) purchases, (iii) business travels and (iv) low carbon science, to propose and analyse the impact of around 60 measures, submitted to consultation meetings and a survey for laboratory members. As a result, a reduction of 7\%/year until 2030 (wrt 2019) was voted, with an ensemble of 7 priority measures, distributed into 7 working groups working towards their realisation. This whole work is associated with a number of actions to maintain motivation and exchanges, while contributing to IRAP's reputation and image as a leading laboratory in the field of transition issues.
%\textbf{Une phrase de ccl points cles}.
%in the environmental transition process. 
%In general, purchases of goods dominate on average by half the carbon footprint of a research institute in France \citep{depaepe2023}. Added to that, astrophysics uses  accounting for about 65\% of its carbon footprint. This together raise the question of the brakes and levers on decarbonisation of research in astrophysics.

\paragraph{Ma Terre en 180 minutes}
In addition to the usual contributed talks and discussion, we organized sessions of the workshop \emph{Ma Terre en 180 min} \citep[``My Earth in 180 min'';][]{gratiot:hal-04126329}, 
%\footnote{\href{https://materre.osug.fr/}{https://materre.osug.fr/}, mostly in French.}, 
a collaborative role play game to build scenarios for reducing the carbon footprint of research, in its \emph{Astronomy \& Planets} game board version. The 2h-long sessions took place during lunch breaks (lunch boxes were provided), from Monday to Thursday. Seven tables were animated, each table gathering 6 people managed by one or two trained animator. In total, 45 attendees of the SF2A conference participated to the MT180 game (10\% of the total on-site participants). Outcomes of this massive session focused on research in astronomy are gathered in \cite{malbet2024}. %\textbf{Une phrase de ccl points cles?}.

%%--------------------
\section{Conclusions}
%%--------------------
In recent years, the French academic system has witnessed an important development around the themes of the environmental transition. The majority of researchers are now aware of the consequences of the environmental crises for the society and for their research. Institutes and research areas have measured and quantified their ecological footprint (in carbon dioxide equivalent). % $\mathtxt{CO}_2\mathtxt{eq}$. 
After awareness and measures, the French academic system, and in particular the A\&A community, is on the verge of committing itself to concrete actions, as shown, for example, by the ubiquity of the environmental constraints in the scientific prospective studies. %short- and long-term

% Optional acknowledgements
% -------------------------
\begin{acknowledgements}
The past and present members of the \emph{Commission Transition Environnementale} would like to warmly thanks the SF2A members of the bureau for their support in officially setting up the commission. 
The S11-SOC would like to thank the members of the SOC and LOC of the SF2A-2024 Conference for accepting the session organisation and for ensuring the smooth running of the session. 
\end{acknowledgements}

\bibliographystyle{aa}  % A&A bibliography style file (aa.bst)
\bibliography{Cantalloube_S11} % your references in file: Yourfile.bib

\begin{thebibliography}{8}
\expandafter\ifx\csname natexlab\endcsname\relax\def\natexlab#1{#1}\fi

\bibitem[{Cantalloube \& No\^us(2024)}]{cantalloube2024a}
Cantalloube, F. \& No\^us, C. 2024, SF2A proceeding

\bibitem[{Fargette(2024)}]{fargette2024}
Fargette, N. 2024, SF2A proceeding

\bibitem[{Gratiot {et~al.}(2023)Gratiot, Klein, Challet, Dangles, Janicot,
  Candelas, Sarret, Panthou, Hingray, Champollion, Montillaud, Bellemain, Marc,
  Bationo, Monnier, Laffont, Foujols, Riffault, Tinel, Mignot, Philippon,
  Dezetter, Caron, Piton, Verney-Carron, Delaballe, Bardet, Nozay-Maurice,
  Loison, Delbart, Anquetin, Immel, Baehr, Malbet, Berni, Delattre, Echevin,
  Petitdidier, Aumont, Michau, Bijon, Vidal, Pinel, Biabiany, Grevesse, Mimeau,
  Biarn{\`e}s, R{\'e}capet, Costes-Thir{\'e}, Poupaud, Barret, Bonnin,
  Mournetas, Tourancheau, Goldman, Bonnet, \&
  Michaud~Soret}]{gratiot:hal-04126329}
Gratiot, N., Klein, J., Challet, M., {et~al.} 2023, {PLOS Sustainability and
  Transformation}, 2, e0000049

\bibitem[{Hardy(2024)}]{hardy2024}
Hardy, A. 2024, Review of Agricultural, Food and Environmental Studies, 105,
  179

\bibitem[{Hennebelle {et~al.}(2024)Hennebelle, Barsuglia, Billebaud, Bouffard,
  Champollion, Grybos, Meheut, Parmentier, \& Petitjean}]{hennebelle2024}
Hennebelle, P., Barsuglia, M., Billebaud, F., {et~al.} 2024, SF2A proceeding

\bibitem[{Malbet {et~al.}(2024)Malbet, Santerne, Milli, Champollion, Lamy,
  Imbaud, Gaunet, Masson, Daré, Gratiot, \& Bellemain}]{malbet2024}
Malbet, F., Santerne, A., Milli, J., {et~al.} 2024, SF2A proceeding
  (hal-04733241)

\bibitem[{Martin {et~al.}(2022)Martin, Brau-Nogu{\'e}, Coriat, Garnier, Hughes,
  Kn{\"o}dlseder, \& Tibaldo}]{martin2022}
Martin, P., Brau-Nogu{\'e}, S., Coriat, M., {et~al.} 2022, Nature Astronomy, 6,
  1219

\bibitem[{Milli {et~al.}(2024)Milli, Boribon, Malbet, Deverchere, Drillat,
  Falque, \& Colas}]{milli2024}
Milli, J., Boribon, M., Malbet, F., {et~al.} 2024, SF2A proceeding
  (arxiv:2410.07280)

\end{thebibliography}

\end{document}